\documentclass[showpacs,twocolumn,pre,floatfix,superscriptaddress]{revtex4}

\usepackage[dvips]{graphicx}
\usepackage[dvips]{color}
\newcommand{\href}[2]{#2}

\usepackage{amsmath}
\usepackage{amssymb}
\usepackage{bm}
\usepackage{times}

\newcommand{\vct}[1]{\mathbf{#1}}
\newcommand{\uvct}[1]{\mathbf{\hat{#1}}}
\newcommand{\tens}[1]{\mathbf{#1}}
\newcommand{\grad}{\bm{\nabla}}

\newcommand{\Weq}{W_{\!\!\text{eq}}}
\newcommand{\HF}{\mathcal{F}}
\newcommand{\HFex}{\mathcal{F}_{\!\!\text{ex}}}
\newcommand{\corr}{\rho^{(2)}}

\newcommand{\correqpsi}{\rho^{(2)}_{\Psi}}

\newcommand{\rhoeq}{\rho_{\text{eq}}}

\begin{document}

\title{A dynamic density functional theory for particles in a flowing solvent}
\date{\today}

\author{Markus Rauscher}
\email{rauscher@mf.mpg.de}
\affiliation{Max-Planck-Institut f\"{u}r
Metallforschung, Heisenbergstr.\ 3, D-70569 Stuttgart, Germany, and \\
Institut f{\"u}r Theoretische und Angewandte Physik, Universit\"{a}t 
Stuttgart,  Pfaffenwaldring 57, D-70569 Stuttgart, Germany}
\author{Alvaro Dom{\'\i}nguez}
\affiliation{F{\'\i}sica Te\'orica, Universidad de Sevilla, Apdo.~1065, E-41080 Sevilla, Spain}
\author{Matthias Kr{\"u}ger}
\affiliation{Max-Planck-Institut f\"{u}r
Metallforschung, Heisenbergstr.\ 3, D-70569 Stuttgart, Germany, and \\
Institut f{\"u}r Theoretische und Angewandte Physik, Universit\"{a}t 
Stuttgart,  Pfaffenwaldring 57, D-70569 Stuttgart, Germany}
\author{Florencia Penna}
\affiliation{Universidad Aut{\'o}noma de Madrid, E-28049 Madrid, Spain}

\begin{abstract}
We present a \textit{dynamic density functional theory} (dDFT) which takes into account the
advection of the particles by a flowing solvent. For potential
flows we can use the same closure as in the absence of solvent
flow. The structure of the resulting advected dDFT suggests that it could be
used for non-potential flows as well. We apply this dDFT to
Brownian particles (e.g., polymer coils) 
in a solvent flowing around a spherical obstacle (e.g., a colloid) and
compare the results with direct simulations of the underlying
Brownian dynamics.
Although numerical limitations do not allow for an accurate quantitative
check of the advected dDFT both show the same qualitative features.
In contrast to previous works which neglected the deformation of the
flow by the obstacle, we find that the bow--wave in the density
distribution of particles in front of the obstacle as well as the
wake behind it are reduced dramatically. 
As a consequence the friction force exerted by the (polymer) particles 
on the colloid can be reduced drastically. 
\end{abstract}


\maketitle

\section{Introduction}\label{sec:intro}

\begin{figure}
\begin{center}
\includegraphics[width=0.49\linewidth]{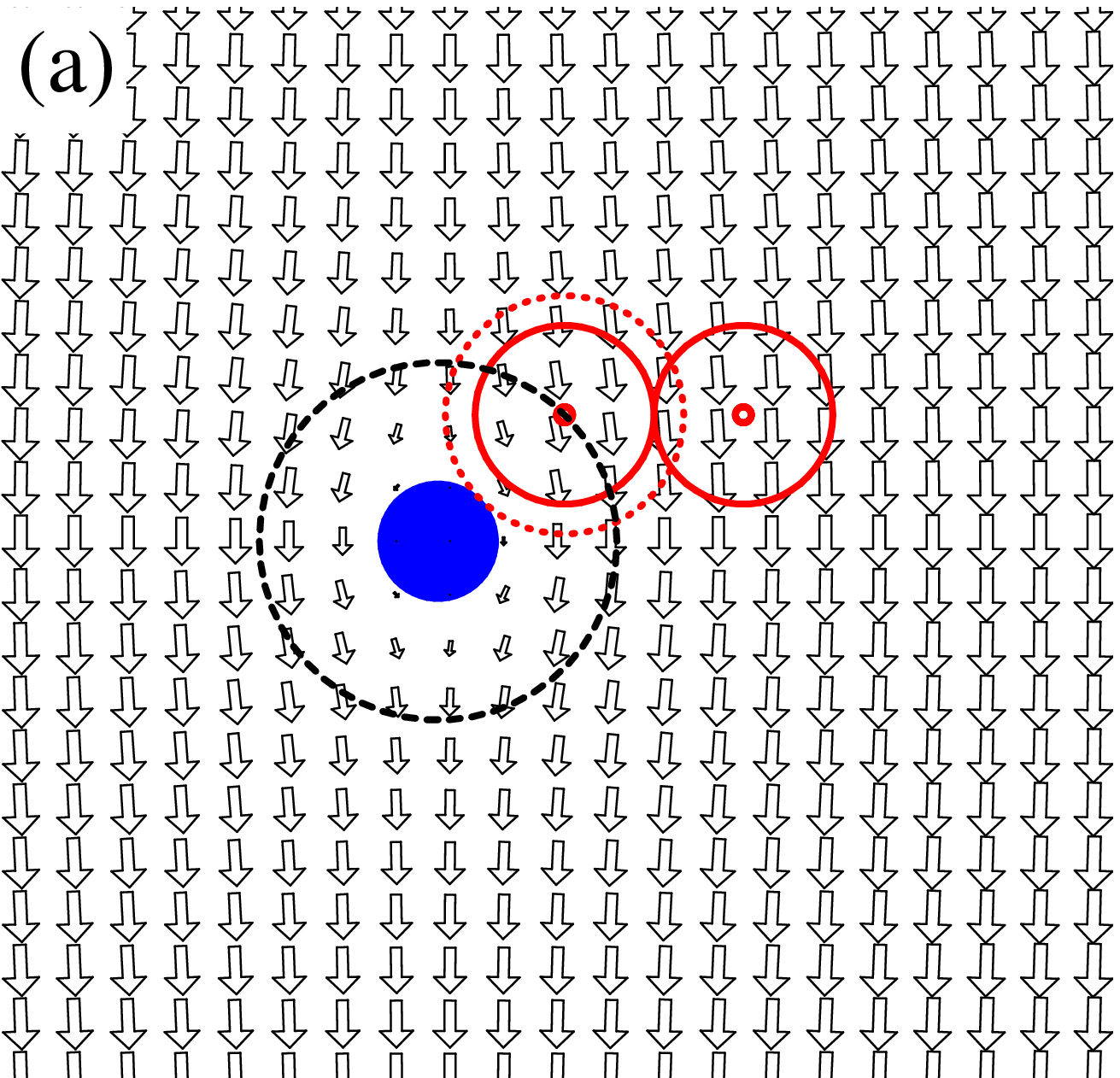}\hfill
\includegraphics[width=0.49\linewidth]{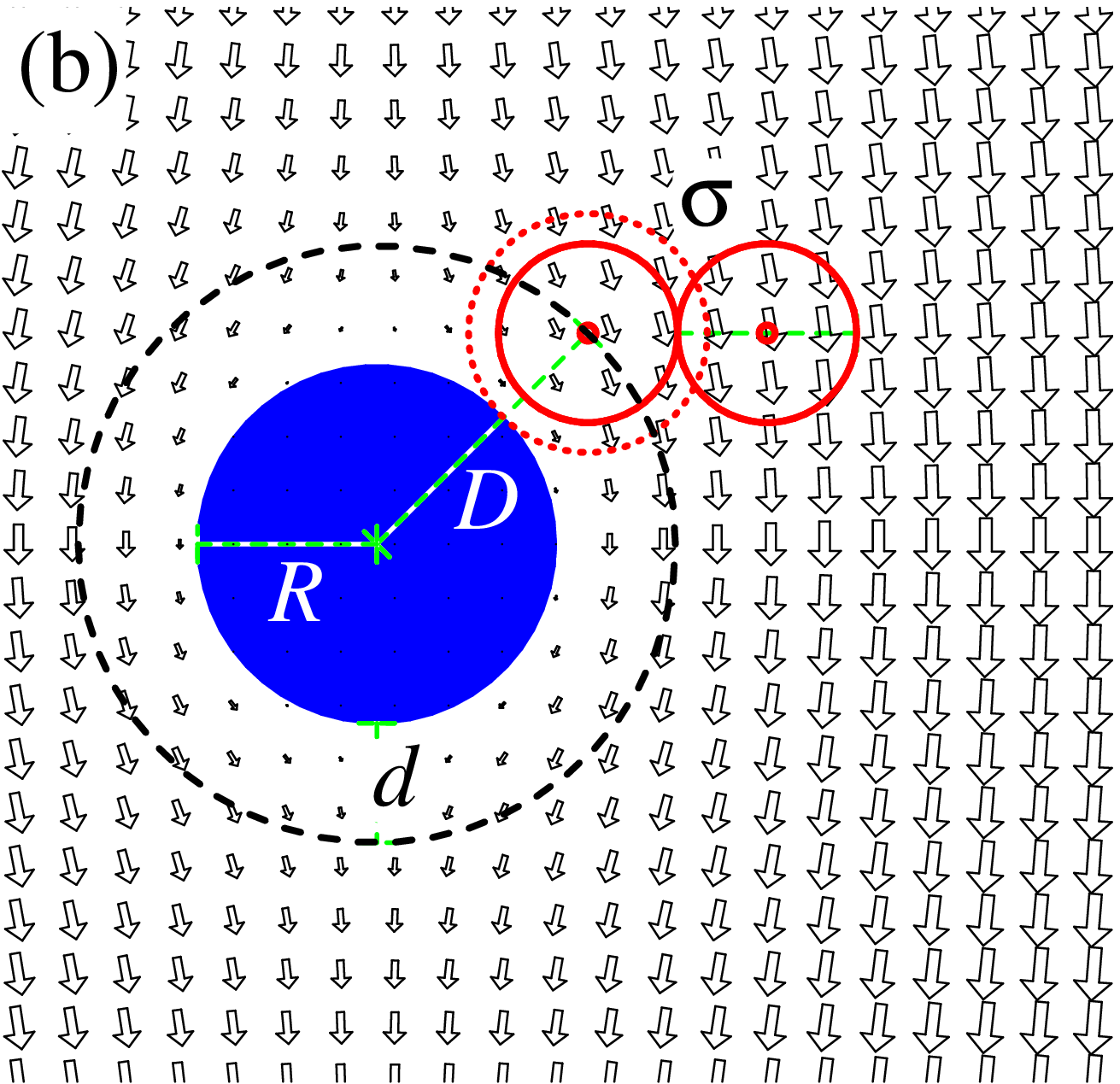}
\end{center}
\caption{\label{fig:flow} Cross section of the flow field given by
Eq.~(\protect\ref{eq:sphereflow}) in a plane parallel to the
direction of motion around (a) a small spherical colloid and (b) a
big one (full circles, radius $R$)\/. The dashed circles
of radius $D$ mark the points of closest approach of solute's centers (point in
the center of open circles) to the colloid\/. The solute's diameter
in its mutual interaction 
is $\sigma$ and, for non-additive mixtures, not necessarily equal
to its diameter $2\,d=2\,(D-R)$ (indicated by the dotted circle) 
in the interaction with the colloid\/.
The component of the flow field normal to the dashed circle is larger for the
small colloid (a) than for the large colloid (b)\/. 
$\sigma$ and $d$ are the same in both figures.
}
\end{figure}

The generalization of classical density functional theory (DFT) to
non-equilibrium states has become a valuable tool to study the
dynamics of 
interacting Brownian particles.  This dynamic
density functional theory (dDFT) for the ensemble averaged density
was proposed recently in \cite{marconi99,marconi00}\/.  Although
hydrodynamics is known to play a crucial role in the dynamics of
suspensions, dDFT has also been used to describe colloidal suspensions
or polymer solutions, in particular 
to investigate the distribution of solute particles 
around a strongly repulsive potential moving through the solution of
particles \cite{penna03b}\/. The intention was to model a
colloidal particle moving through a polymer solution.  A similar
model has been used to study the depletion interaction
between two colloidal particles moving through a polymer solution
\cite{dzubiella03b,krueger07}\/. In all these studies the hydrodynamic flow of the solvent
around the colloid was neglected and the solvent effectively passed through
the colloid. A real colloid would displace the solvent as it
moves, as shown for the case of a small and a large colloid in
Fig.~\ref{fig:flow}\/. For a spherical colloidal particle of radius $R$
dragged through an unbounded incompressible viscous Newtonian solvent 
with velocity $\vct{c}$ at low Reynolds number, the flow field
$\vct{u}(\vct{r})$ (in a frame of reference comoving with the
colloid) is given by the solution of the Stokes equation
\cite{landaulifschitzVI},
\begin{equation}
\label{eq:sphereflow}
\vct{u}(\vct{r}) = \frac{3\,R}{4\,r}\,\left(1+\frac{R^2}{3\,r^2}\right)\,
\vct{c} + \frac{3\,R}{4\,r^3}\,\vct{r}\,(\vct{r}\cdot
\vct{c})\,\left(1-\frac{R^2}{r^2}\right) -\vct{c}.
\end{equation}
For large distances from the colloid, $r\gg R$, the flow field is well
approximated by $\vct{u}(\vct{r})=-\vct{c}$\/. For solute particles
(e.g., polymers or other colloids) which only feel this far field, the
model presented in \cite{penna03b,dzubiella03b,krueger07} is a reasonable
approximation. This is the case for large solute particles with a
radius $d \gg R$\/. Their centers can approach the dragged 
colloidal particle only up to a
distance $D=R+d \approx d$, see Fig.~\ref{fig:flow}(a),
and thus will feel a flow field $\vct{u}(\vct{r})\approx -\vct{c}$, 
if one neglects the additional effect of the solute
particles on the solvent flow, in other words, the hydrodynamic
interaction between the solute particle and the colloid is
neglected. This is a reasonable approximation for polymer coils 
but certainly a bad one for solid solute particles.
But small solute particles of radius $d\ll R$ can get much closer to
the colloid and feel the 
distortion of the solvent velocity field as illustrated in
Fig.~\ref{fig:flow}(b)\/. The solute particles will be deviated from the
colloid by the flow field and this will reduce significantly the
bow--wave effect in front of the colloid presented in
\cite{penna03b} and the strenght of the non-equilibrium
depletion force discussed in \cite{dzubiella03b,krueger07}\/. Extremely
small solute particles would not show this effect at all since they
would behave like solvent molecules. In this limit, however, the
basis of the theory discussed here, i.e, the description of the
solute particles as overdamped Brownian particles, is no longer
valid because it is based on a separation of the length and time 
scales associated with the solvent molecules and the solute particles.

In this paper we present a generalization of the dDFT 
derived in \cite{marconi99,marconi00} to the
case of Brownian solvent particles advected by a flow, thereby
incorporating some aspects of the hydrodynamics of the solvent
into the theory. 
However, we do not model 
hydrodynamic interactions between the solute particles as well as the
back-reaction of the solute particles 
on the flow field, e.g., by a
concentration dependent viscosity, or by a reduced mobility
of the solute particles in the vicinity of the colloid. 
However, the latter can be included in a straightforward manner as 
we discuss in the
conclusions in Sec.~\ref{sec:conclusions}\/. In the following
Sec.~\ref{sec:ddft} we derive the advected dynamic density functional 
theory using the method described in \cite{archer04a,archer04b}\/. 
In Sec.~\ref{sec:examples} we discuss two sample cases, namely ideal
solute particles and Gaussian solute particles which stress the importance of
taking into account the solvent flow. 

\section{Advected \lowercase{d}DFT} \label{sec:ddft}

We start with the Langevin equation of an ensemble of $N$ advected 
interacting Brownian particles confined to a finite volume $\mathcal{V}$ 
in the overdamped limit,
\begin{equation}
\label{eq:brownian}
\frac{d \vct{r}_i}{d t} = \vct{u}(\vct{r}_i)-
\Gamma\,\grad_i\left[U(\vct{r}_i)+
\sum\limits_{j=1}^N V(|\vct{r}_i - \vct{r}_j|) 
\right]
+ \bm{\eta}_i(t),
\end{equation}
with the pair interaction potential between the particles
$V(r)$ and an external potential $U(\vct{r})$\/. Both the external
potential $U(\vct{r})$ and the flow field $\vct{u}(\vct{r})$ can
depend on time. 
For clarity of notation we will not make this
dependence explicit in the equations. 
The flow field is not necessarily
divergence free (i.e., the solvent can be compressible)\/.
$\grad_i$ denotes the gradient with
respect to $\vct{r}_i$\/. We approximate the noise generated by the
thermal motion of the solvent particles by a Wiener process, 
\begin{align}
\label{eq:noiseaverage}
\langle \bm{\eta}_i(t) \rangle &=\bm{0} \quad\quad\quad\text{and}\\
\label{eq:noisecorrelator}
\langle \eta_i^\alpha(s)\,\eta_j^\beta(t)\rangle
&=2\,T\,\Gamma\,\delta_{ij}\,\delta_{\alpha\beta}\,\delta(t-s),
\end{align}
with the temperature $T$ measured in units of energy (setting $k_{\!B}=1$)
and the mobility coefficient $\Gamma>0$\/. The mobility coefficient has to
appear in the correlation in Eq.~(\ref{eq:noisecorrelator}) in order to
fulfill the fluctuation--dissipation theorem and to get the correct
equilibrium distribution for $\vct{u}(\vct{r}) =\bm{0}$\/. The boundaries
of $\mathcal{V}$ are impermeable for the particles or periodic (or a
mixture of both), and therefore the number of particles is conserved.

The Fokker-Planck equation corresponding to the Langevin equation
(\ref{eq:brownian}) gives the time evolution of the probability density 
$W(\vct{r}_1,\dots,\vct{r}_N,t)$ for
finding the particles at time $t$ at the positions 
$\vct{r}_1,\dots,\vct{r}_N$ \cite{gardiner83,risken84},
\begin{multline}
\label{eq:fokkerplanck}
\frac{\partial W}{\partial t} = -
\sum\limits_{i=1}^N \grad_i \cdot \Bigg\{\Gamma\Bigg[
\frac{\vct{u}(\vct{r}_i)}{\Gamma}-\grad_i U(\vct{r}_i)\\
-\sum\limits_{j=1}^N \grad_i V(|\vct{r}_i - \vct{r}_j|)
-T\,\grad_i\Bigg]\,W\Bigg\}.
\end{multline}
For a potential flow, the velocity field can be written as the
gradient of a scalar field, $\vct{u}(\vct{r})= - \Gamma\,\grad
\Phi(\vct{r})$ \footnote{Note that $\Phi$ has to be uniquely defined up to a constant, also for periodic boundaries of
$\mathcal{V}$\/. For example, for the uniform flow field
$\vct{u}=(u_x,0,0)$ with periodic boundaries in $x$-direction this is not
the case.},
such that the external potential and the effect of the flow
field can be combined into a modified external potential $U^*(\vct{r}) =
U(\vct{r})+\Phi(\vct{r})$\/. 
If $U^*$ is time--independent, one can find a stationary probability
density $\Weq^*(\vct{r}_1,\dots, \vct{r}_N)$ which fulfills the detailed 
balance condition for Eq.~(\ref{eq:fokkerplanck}), i.e., 
the term in curly brackets is zero for each $i=1,\dots,N$, 
\begin{equation}
  \label{eq:detailedbalance}
  \Bigg[ \grad_i U^*(\vct{r}_i)\\
  +\sum\limits_{j=1}^N \grad_i V(|\vct{r}_i - \vct{r}_j|)
  +T\,\grad_i\Bigg]\,\Weq^* = \bm{0}.
\end{equation}
The solution is
\begin{equation}
\label{eq:eqdistribution}
\Weq^*(\vct{r}_1,\dots, \vct{r}_N)=\frac{1}{\mathcal{Z}^*}\,
\mathrm{e}^{-\frac{1}{T}\,
\sum\limits_{i=1}^{N}\left[
U^*(\vct{r}_i)+\sum\limits_{j=1}^N V(|\vct{r}_i - \vct{r}_j|)
\right]},
\end{equation}
normalized with the sum of states $\mathcal{Z}^*$ such that 
\begin{equation}
\iint\limits_{\mathcal{V}^N} d^3r_1\dots d^3r_N \, \Weq^*(\vct{r}_1,\dots, \vct{r}_N)=1.
\end{equation}
For such a situation the whole apparatus of equilibrium statistical
mechanics can be used in order to calculate expectation values and
correlations in a stationary non-equilibrium situation. But this is
restricted to cases when the detailed balance condition holds, which
implies necessarily a potential flow.
Even in the Stokes flow~(\ref{eq:sphereflow}) or in a
simple shear flow (e.g., in Couette or Poiseuille flow) this is not 
true because $\grad\times \vct{u} \ne \bm{0}$\/. 
For flows with a finite vorticity
there is no detailed balance in a strict sense, see
\cite[Eq.~(5.3.4(c))]{gardiner83}\/. 

From Eq.~(\ref{eq:fokkerplanck}) we can calculate the time evolution of the
noise averaged particle density $\rho(\vct{r},t)$, namely, the expectation 
value of the density operator $\hat\rho(\vct{r},t) = \sum_{i=1}^N
\delta(\vct{r}-\vct{r}_i(t))$, 
\begin{align}
\label{eq:hierarchy}
\frac{\partial \rho}{\partial t} +\grad\cdot(\rho\,\vct{u}) &= 
\grad\cdot\Gamma\,\bigg[
\rho\,\grad U + T\,\grad\rho \\
&+ \grad\,\int\limits_{\mathcal{V}} d^3r'\,V(|\vct{r}-\vct{r}'|)
\corr(\vct{r},\vct{r}',t)
\bigg], \nonumber
\end{align}
with the mean density 
\begin{equation}
\rho(\vct{r},t)= N\,\iint\limits_{\mathcal{V}^{N-1}} d^3r_2\dots
d^3r_N\,W(\vct{r},\vct{r}_2,\dots,\vct{r}_N,t),
\end{equation}
and the non-equilibrium density-density correlation function 
\begin{multline}
\label{eq:corrfunc}
\corr(\vct{r},\vct{r}',t) = N\,(N-1)\\
\times\iint\limits_{\mathcal{V}^{N-2}} d^3r_3\dots d^3r_N\,
W(\vct{r},\vct{r}',\vct{r}_3,\dots,\vct{r}_N,t).
\end{multline}
Eq.~(\ref{eq:hierarchy}) is the starting point of a hierarchy of $N$ 
evolution equations which connect the time derivative of the $n$-point
density correlation function to the $n+1$-point density correlation
function, similar to the BBGKY hierarchy for deterministic systems with
inertia or the BGY hierarchy for equilibrium correlation functions.

In order to find a closed equation for the time evolution of
$\rho(\vct{r},t)$ we approximate the interaction term in Eq.~(\ref{eq:hierarchy}) by its value in an equilibrium system with the
same interaction potential $V(r)$\/. 
Let us first restrict our considerations to the case that detailed balance holds (so that, in particular, $\vct{u} = - \Gamma\,\grad\Phi$).
%
We modify our system of Brownian particles by applying an external
potential $\Psi(\vct{r})$ so as to
create a system whose equilibrium density distribution is $\rhoeq^\Psi(\vct{r})= \rho(\vct{r},t)$\/. This new potential $\Psi(\vct{r})$ depends on $\rho(\vct{r},t)$ and will
be different for each $t$ as long as $\rho(\vct{r},t)$ is not stationary.
The Fokker-Planck equation for the modified system will be
Eq.~(\ref{eq:fokkerplanck}) but with
$U(\vct{r})$ replaced by $U(\vct{r})+\Psi(\vct{r})$\/.
The equilibrium probability density $\Weq^\Psi$
of the modified system 
is given by Eq.~(\ref{eq:eqdistribution}) but with
$U^*(\vct{r})$ replaced by 
$U^*(\vct{r})+\Psi(\vct{r})$. If we integrate
the detailed balance condition~(\ref{eq:detailedbalance}) for $\Weq^\Psi$ over $N-1$ positions, we get
\begin{multline}
\label{eq:detailedbalancerho}
\vct{u}\,\rhoeq^\Psi = \Gamma\,\bigg[
\rhoeq^\Psi\,\grad \left(U +\Psi\right) +T\,\grad\rhoeq^\Psi 
\\+\grad\int\limits_{\mathcal{V}} d^3r'\,
V(|\vct{r}-\vct{r}'|)\,\correqpsi(\vct{r},\vct{r}')\bigg],
\end{multline}
with the equilibrium pair correlation function
$\correqpsi(\vct{r},\vct{r}')$ for the modified system in the external
potential $\Psi$\/.
From equilibrium density functional theory one knows that the equilibrium
density distribution in the grand canonical ensemble  is the minimum of the
grand canonical functional 
\begin{multline}
\label{eq:denstiyfunctional}
\Omega[\rho] = 
\HFex[\rho]\\
+\int\limits_{\mathcal{V}} d^3r\,\left\{
T\,\rho\left[\ln\left(\rho\,\Lambda^3\right)-1\right]
+\left(U_{\text{ext}}-\mu\right)\,\rho\right\} ,
\end{multline}
with the thermal wavelength $\Lambda$, the chemical potential $\mu$ and the
sum of all external potentials $U_{\text{ext}}=U+\Phi+\Psi$\/. The excess free energy
$\HFex[\rho]$ summarizes the effect of the particle interactions and it is
not known exactly in general. 
We take the gradient of the
Euler-Lagrange equation following from the functional~(\ref{eq:denstiyfunctional}): since in thermal equilibrium the chemical potential
is constant across the whole system we get
\begin{equation}
\label{eq:gradeulerlagrange}
\frac{\vct{u}}{\Gamma} = \grad(U +\Psi)
+\frac{T}{\rhoeq^\Psi}\,\grad\rhoeq^\Psi +
\grad\left.\frac{\delta\HFex[\rho]}{\delta\rho}\right|_{\rhoeq^\Psi}.
\end{equation}
If we compare Eq.~(\ref{eq:detailedbalancerho}) with
(\ref{eq:gradeulerlagrange}) we can see that
\begin{equation}
\label{eq:closure}
\grad\int\limits_{\mathcal{V}} d^3r'\,
V(|\vct{r}-\vct{r}'|)\,\correqpsi(\vct{r},\vct{r}')
=\rhoeq^\Psi\,
\grad\left.\frac{\delta\HFex[\rho]}{\delta\rho}\right|_{\rhoeq^\Psi}.
\end{equation} Note that the right hand side does not depend on the
velocity potential $\Phi$ while the dependence on $\Psi$ enters
only through $\rhoeq^\Psi$\/. We will use Eq.~(\ref{eq:closure})
as a closure to the hierarchy of equations starting with
Eq.~(\ref{eq:hierarchy})\/: Hereby we assume that the density
correlations at time $t$ in the non-equilibrium system with mean
density $\rho(\vct{r},t)$ are the same as in an equilibrium system
with the additional potential $\Psi$ and with equilibrium mean
density $\rhoeq^\Psi(\vct{r})=\rho(\vct{r},t)$\/. We then get
\begin{equation}
\label{eq:dDFT}
\frac{\partial \rho}{\partial t}+\grad\cdot(\rho\,\vct{u})=
\grad\cdot\left(\Gamma\,\rho\,\grad\frac{\delta \HF[\rho]}{\delta\rho}\right),
\end{equation}
with the free energy functional 
\begin{equation}
\label{eq:freeenergy}
\HF[\rho]=\HFex[\rho]+\int\limits_{\mathcal{V}}
d^3r\,\left\{T\,\rho\left[\ln\left(\rho\,\Lambda^3\right)-1\right]
+\rho\, U \right\}.
\end{equation}

In thermodynamic equilibrium for $\vct{u}=0$ and time--independent
$U$, the equilibrium density distribution given by
$\mu=\left.\frac{\delta\HF[\rho]}{\delta\rho}\right|_{\rhoeq}$ is a
stationary solution of Eq.~(\ref{eq:dDFT})\/. An H-theorem
\begin{equation}
\frac{\partial}{\partial t}\int\limits_{\mathcal{V}} d^3r \, \HF[\rho]= 
-\int\limits_{\mathcal{V}} d^3r\, \Gamma\,\rho\,\left(\grad \frac{\delta \HF[\rho]}{\delta\rho}
\right)^2 \le 0
\end{equation}
guarantees that the time evolution actually converges to the
equilibrium distribution. (The dynamics in Eq.~(\ref{eq:brownian})
together with the boundary conditions taken for ${\cal V}$ imply
that the particle current through the system boundaries is zero and
therefore surface terms from partial integration vanish.)
The final chemical potential is then determined by the conserved
number of particles in the system. 
As discussed above, a system in a potential flow corresponds to an
equilibrium system with a modified external potential
$U^*=U+\Psi$\/.  Eq.~(\ref{eq:dDFT}) can then be written in the
form
\begin{equation}
\frac{\partial \rho}{\partial t} =
\grad\left(\Gamma\,\rho\,\grad\frac{\delta
\HF^*[\rho]}{\delta\rho}\right),
\end{equation}
with the modified free energy functional $\HF^*[\rho] = \HF[\rho] +
\int\limits_{\mathcal{V}} d^3r\,\Phi\,\rho$\/. Thus
we have an H-theorem for $\HF^*[\rho]$
instead of $\HF[\rho]$ and the ``equilibrium state'' is determined by
$\frac{\delta\HF^*[\rho]}{\delta\rho} = \mu^*$\/.

The right hand side of Eq.~(\ref{eq:closure}) is completely
independent of the flow field and one could be tempted to use it as
a closure to Eq.~(\ref{eq:hierarchy}) for the most general case
(i.e., non-potential flows). However, this would mean approximating
density correlations in a driven non-equilibrium system where
detailed balance cannot be achieved
by thermal equilibrium correlations. While 
we could argue that close to equilibrium Eq.~(\ref{eq:closure}) may
be a reasonable approximation if detailed balance still holds,
there is no such argument for the most general case that detailed
balance is violated. 
The study of Ref.~\cite{penna03b} addresses a situation where the
approximation is found to be good although there is no detailed balance 
since a net particle current is driven through the system.
%
In the next Section we consider some examples with the purpose of
assessing (i) the effect of a more realistic Stokes flow as
discussed in the Introduction, and (ii) the validity of the
approximate Eq.~(\ref{eq:dDFT}) for a non-potential flow.

\section{Examples} \label{sec:examples}

As an example we study a solution of polymers (radius $d$,
density $\rho_0$) in an incompressible Newtonian solvent 
flowing around a spherical colloidal particle
(radius $R$) in a stationary situation\/. We model the polymer
coils as point--like particles from the point of view of the
solvent, but with a finite interaction range $\sigma$ concerning
other polymer coils and $D=d+R$ concerning the colloidal particle,
the interaction with the latter being of hard--wall type, see
Fig.~\ref{fig:flow}\/.
The velocity field of the solvent is given by the Stokes flow,
Eq.~(\ref{eq:sphereflow}), and we choose $\vct{c}=c\,\uvct{e}_z$\/.
Measuring lengths in terms of $D$,
the dimensionless parameters determining the system are the P{\'e}clet
number $c^\ast = c\,D/(\Gamma\,T)$, the colloid radius $R^\ast = R/D$, and
the polymer's mutual interaction range $\sigma/D$\/. 

\subsection{Ideal polymers}
\label{sec:ideal}

For ideal solute particles in an incompressible solvent with $U=0$, 
the stationary condition
$\frac{\partial\rho}{\partial t}=0$ from Eq.~(\ref{eq:dDFT}) reads 
\begin{equation}
\label{eq:ideal}
\vct{u}\cdot\grad\rho =\Gamma\,T\,\Delta \rho ,
\end{equation}
where $\Gamma\,T$ is 
the 
diffusion constant of the solute particles.
The hard interaction with the colloid is written as a boundary
condition for the current density of the solute particles
$\vct{j}=\vct{u}\,\rho-\Gamma\,T\,\grad\rho$  at $r/D=1$, 
\begin{equation}
\label{eq:BC}
\left.\left(\uvct{e}_r\cdot\vct{j}\right)\right|_{r/D=1}=0.
\end{equation}
We expand the density field $\rho(\vct{r})$ in spherical harmonics up
to order $N$ and obtain a system of $N+1$ ordinary differential
equations for the $|\vct{r}|$-dependent coefficients which we
solve numerically with AUTO~2000
\footnote{http://sourceforge.net/projects/auto2000/}\/. AUTO~2000 is a
software which solves autonomous boundary value problems for
systems of ordinary differential equations by
continuation, i.e., by starting from a known solution for a specific set of
problem parameters (for $c=0$ we have $\rho=\rho_0$)
and changing parameters (in
our case $c$) continuously until the desired value is reached. 
\begin{figure}
\begin{center}
%
%
%
%
\includegraphics[width=0.49\linewidth]{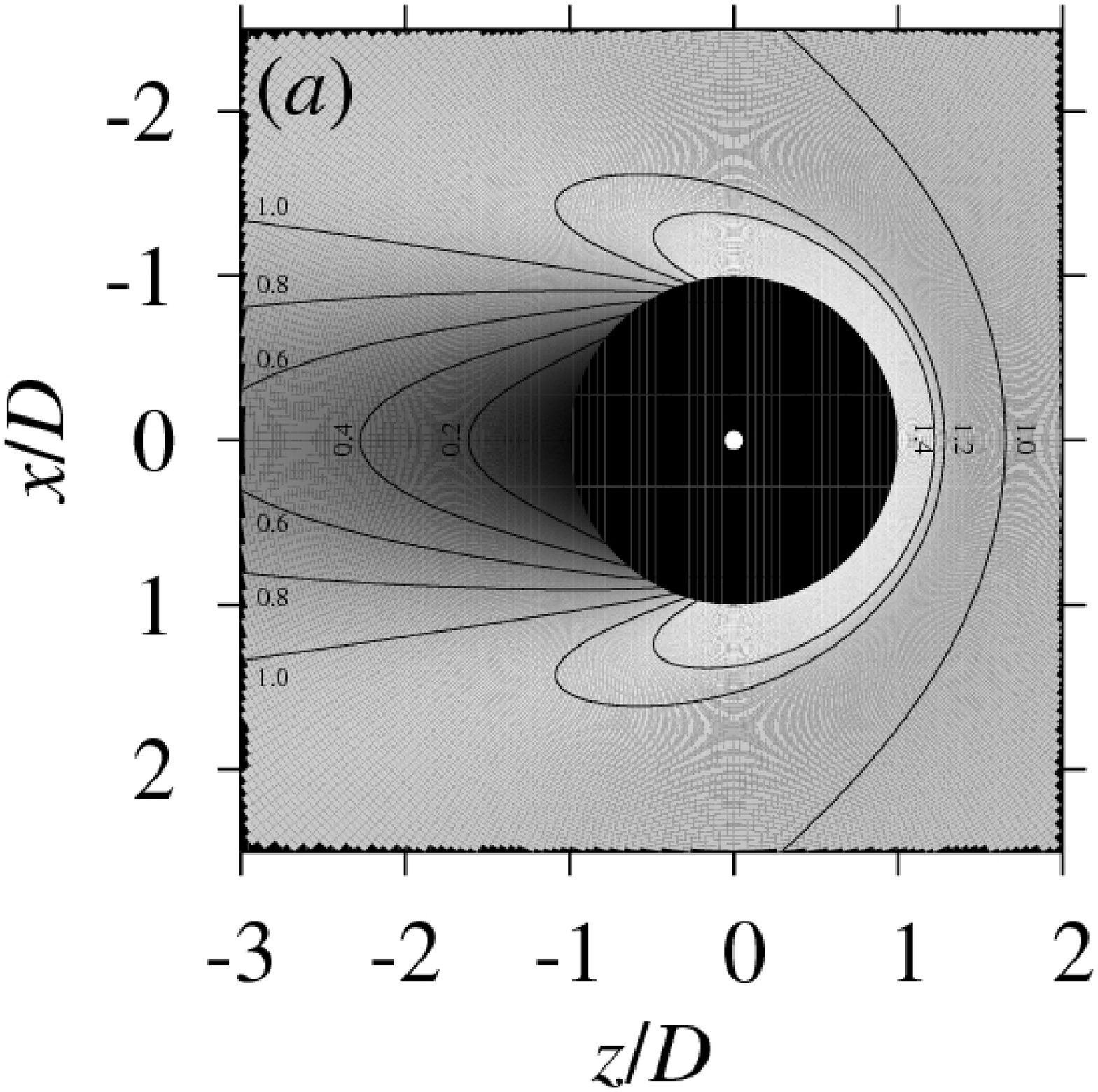}
\includegraphics[width=0.49\linewidth]{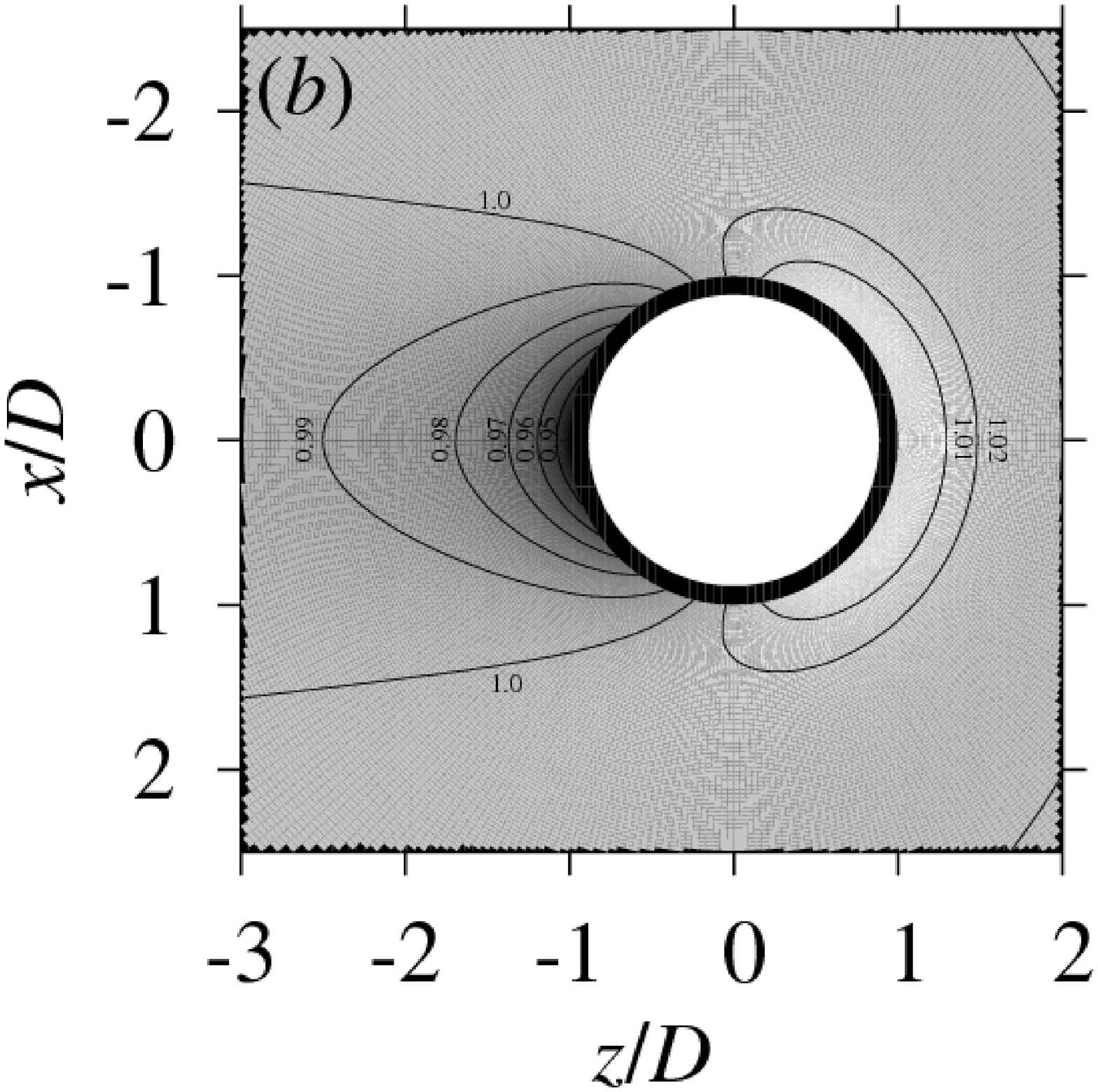}
\end{center}
\caption{\label{fig:cont} Contour plots of the density of ideal
polymers for a flow velocity $c^\ast=10$\/. 
The white circle at the origin is the colloidal particle with radius $R$, 
the black circle is the annulus of thickness $d$ and outer diameter
$D$ which is unaccessible to 
the polymer centers due to the hard-wall interaction, 
c.f. Fig.~\protect\ref{fig:flow}\/.
$(a)$ corresponds to a uniform
flow $\vct{u}(\vct{r})=-c\,\uvct{e}_z$ (i.e., to $R^\ast=0$).
The maximum density in front of the colloid is
$\rho(\vct{r})/\rho_0=6.31$\/. $(b)$ corresponds to
$R^\ast =0.9$\/.
The bow effect is reduced drastically. 
The maximum density in front of the colloid is
$\rho(\vct{r})/\rho_0=1.05$\/.}
\end{figure}

\begin{figure}
\begin{center}
\includegraphics[width=0.49\linewidth]{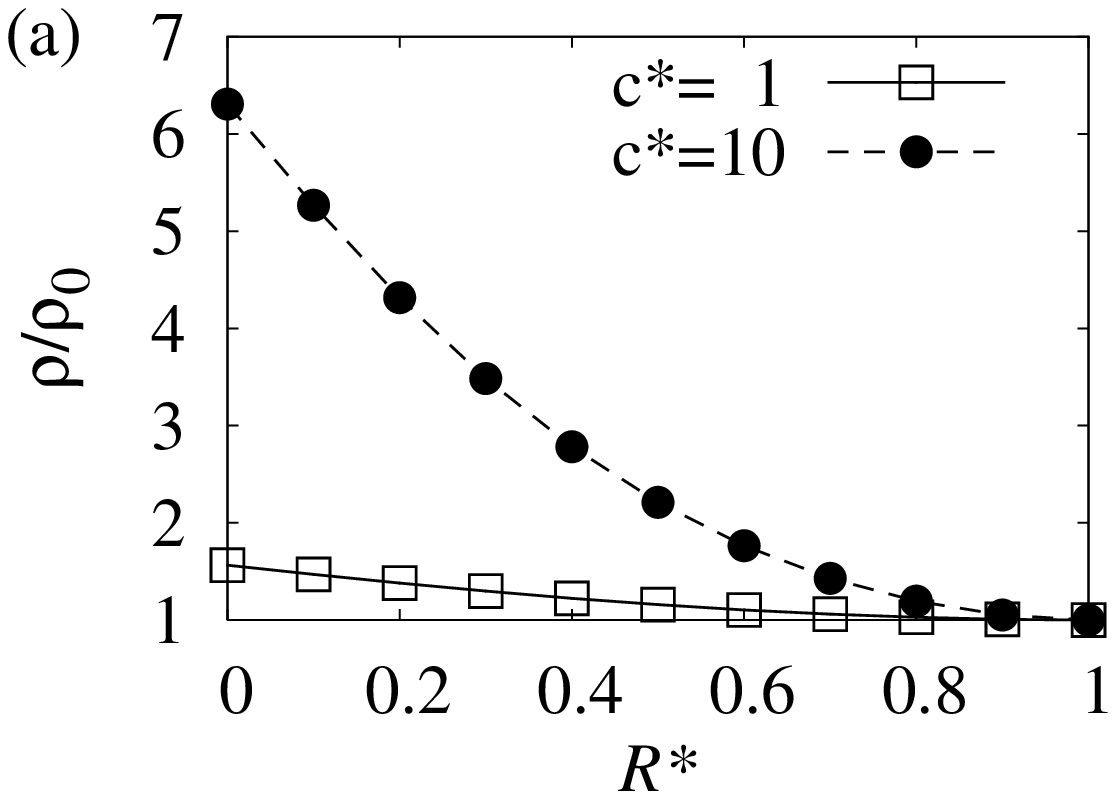}\hfill
\includegraphics[width=0.49\linewidth]{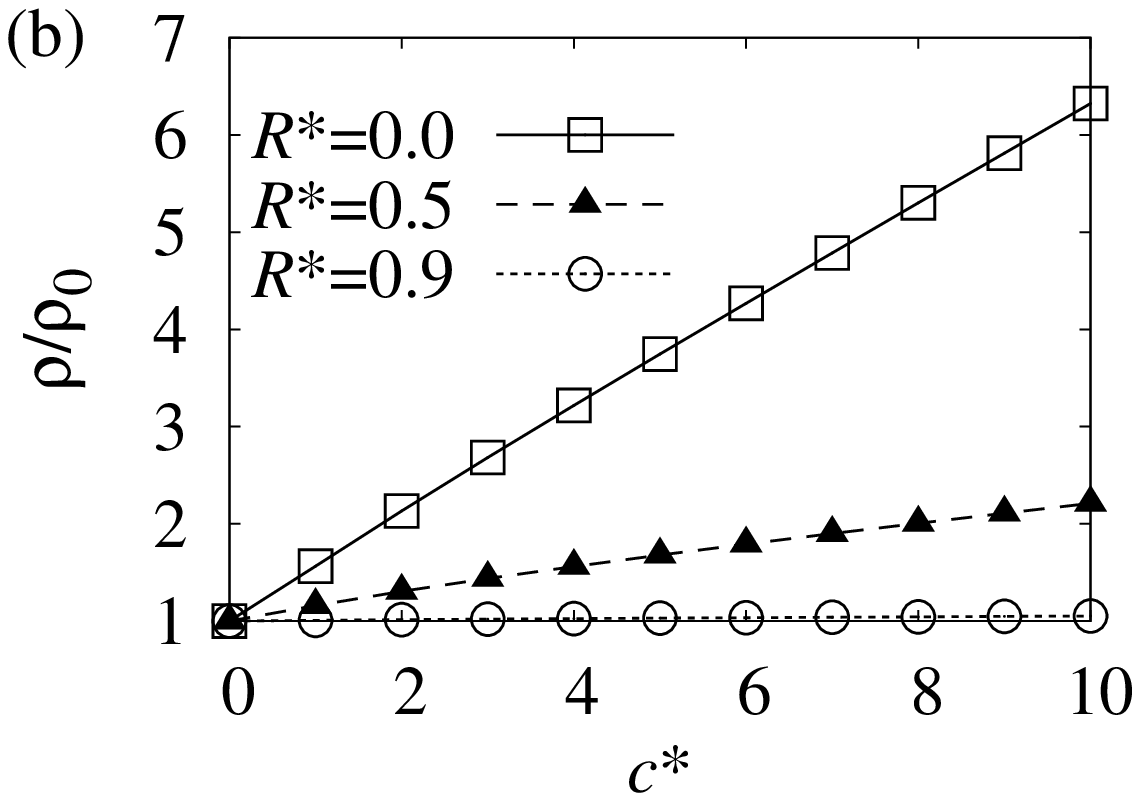}
\end{center}
\caption{\label{fig:plots} Density of ideal polymers at the point
$x=y=0$, $z/D=1$, i.e., right in front of the forbidden zone around
the colloid. (a)
shows $\rho(0,0,D)$ as a function of the colloid size for
$c^\ast=1$ and $c^\ast=10$\/. (b) shows $\rho(0,0,D)$ 
as a function of $c^\ast$ for 
different values of $R^\ast$\/.}
\end{figure}

As demonstrated for $R^\ast=0$ and $R^\ast=0.9$ in
Figs.~\ref{fig:cont}(a) and (b), respectively,
the bow wave effect is large
when the colloid is small compared to the polymers and does
not distort too much the flow, reaching its maximum as $R^\ast\to 0$.
This is the case investigated in \cite{penna03b}\/. When the colloid
is large compared to the polymers, the bow wave effect is
small. In the limit $d/R\to 0$, the effect vanishes completely
since the polymers behave like solvent molecules.
Figs.~\ref{fig:plots}(a) and (b) shows the 
density right in front of the colloid as a function of $R^\ast$ and
of $c^\ast$, respectively. The density of ideal solute particles
scales almost linearly with the velocity $c^\ast$\/.

\subsection{Gaussian polymers}
\label{sec:gauss}

Here we address the case of interacting polymers. We consider the same
polymer--polymer interaction potential studied in
Ref.~\cite{penna03b}, namely
\begin{equation}
  V(r) = T \exp{[-(r/\sigma)^2]} .
\end{equation}
The interaction of the polymers with the colloidal particle is
modelled as an external potential of the form
\begin{equation}
  U(r) = 10\, T \exp{[-(r/a)^6]} .
\end{equation}
This potential rises steeply up to $10\,T$, thus resembling a hard wall.
The length $a$ must be related to the radius of the forbidden zone
$D$ around the colloidal particle.
We conventionally set the value of $a$
by the condition $U(D) = 2\,T$, giving $a\approx 0.924\,D$\/.
We take $\sigma=2 d$ (i.e., additive mixture of polymers and colloidal
particle) and $R=1.7\,\sigma$, leading to $R^\ast \approx 0.77$,
$\sigma/D\approx 0.46$.
Finally, we also considered the choice $R^\ast =0$, $\sigma/D\approx 0.46$,
which represents a hard particle that does not distort the uniform
flow ($R=0$ in Eq.~(\ref{eq:sphereflow})) in a non--additive mixture
($\sigma \neq 2 d$): this was the model addressed in
Ref.~\cite{penna03b}\/.

We ran Brownian dynamics (BD) simulations of this system for two
values of the flow velocity corresponding to the polymer P\'eclet
numbers $(\sigma/D) c^\ast=1$ and $10$ studied in Ref.~\cite{penna03b}
(i.e., $c^\ast\approx 2.2$ and $22$). We considered a colloidal
particle at the center of a box of dimensions $L_x=L_y=12\,\sigma$
and $L_z=24\,\sigma$ 
with periodic boundary conditions. The box contained $N=3456$ polymers,
corresponding to a mean polymer number density $\rho_0\, \sigma^3 =
1$\/. 
We took a timestep of 
$0.003\, \sigma^2 \Gamma/T$ 
for the discretized
Langevin dynamics. The system was allowed to relax for $10^5$
timesteps, after which collection of data was carried out during
$10^6$ timesteps.
Even though the simulated system is finite we used the analytically known
flow field around a sphere in an infinite medium,
Eq.~(\ref{eq:sphereflow})\/. 
The error due to the truncation of this flow by the boundary of the
simulation box is largest (about 20\%) at the midplane of the colloid
($z=0$)\/.
This introduces effectively a discontinuity in the flow velocity field
at the boundary which we discuss later.

We also solved numerically the dDFT in the random phase approximation
(a mean--field model), i.e., with
\begin{equation}
  \HFex[\rho] = \frac{1}{2} \iint\limits_{\mathcal{V}^2} d^3r \, d^3r' \; 
  V(|\vct{r} - \vct{r}'|) \rho(\vct{r}) \rho(\vct{r}') .
\end{equation}
The time evolution given by Eq.~(\ref{eq:dDFT}) of an initially
homogeneous density was solved 
in cylindrical coordinates
on a grid spanning the domain $-120<z/\sigma<24$, $0\leq
r_\perp/\sigma <60$, 
where $r_\perp=\sqrt{x^2+y^2}$. The grid constant
was $0.0125\,\sigma$ 
near the colloid, i.e., for $|z|, r_\perp < 6\,\sigma$,
and $0.1\,\sigma$ 
in the rest of the domain. 
For details on the numerical
procedure see \cite{penna03b}\/. The boundary condition at the
domain border was $\rho=\rho_0$  
also in this case we used the flow field given by
Eq.~(\ref{eq:sphereflow})\/.  The error introduced here is smaller
than in the BD simulations since the integration domain for the
dDFT is larger than the BD simulation box.

\begin{figure}
\begin{center}
\includegraphics[width=1\linewidth]{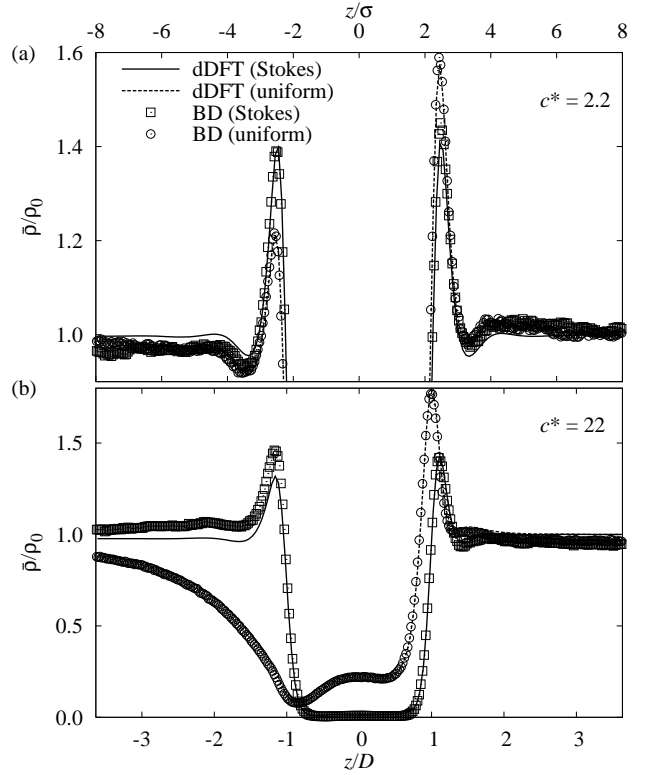}
\end{center}
\caption{ Plots of $\bar{\rho}(z)$ defined in Eq.~(\protect\ref{eq:brho})
  as provided
  by the numerical solution of dDFT (lines) and as measured in BD
  simulations (symbols)\/.
  (a) corresponds to a velocity $c^\ast=2.2$ and (b) to
  $c^\ast=22$\/. In each plot the results for both uniform flow
  ($R^*=0$) and Stokes flow ($R^*=0.77$) are presented.}
\label{fig:BDvsdDFT}
\end{figure}

Fig.~\ref{fig:BDvsdDFT} presents 
the density field $\bar{\rho}(z)$, spatially averaged over thin disks of
radius $\sigma$ 
and thickness $2\,\Delta z=0.05\,\sigma$ 
centered at the $z$-axis, i.e.,
\begin{equation}
  \label{eq:brho}
  \bar{\rho} (z) := \frac{1}{\sigma^2 \Delta z} 
  \int\limits_{z-\Delta z}^{z+\Delta z} dz'
  \int\limits_0^\sigma dr_\perp
  r_\perp \, \rho(r_\perp, z') .
\end{equation}
The results of both the BD simulations and the dDFT illustrate the
dramatic effect of advection by the Stokes flow~(\ref{eq:sphereflow}).
In particular, at the higher velocity $c^\ast=22$ and $R^\ast=0$ (uniform
flow) there is a marked accumulation of polymers in front of the
particle and a strong depletion behind it which are hardly observable
for $R^\ast=0.77$\/. In general, the effect of the Stokes flow is to
weaken 
the influence of the 
colloid
on the density profile,
as the polymers tend to be advected by the stream and to travel
around the particle. Actually, for $c^\ast=2.2$ (and smaller) 
the deformation by the Stokes flow
is so tiny that the dDFT profile $\bar{\rho}(z)$ in 
Fig.~\ref{fig:BDvsdDFT}(a) is 
indistinguishable from the equilibrium profile (i.e.,
$c^\ast=0$)\/.

\begin{figure}
\begin{center}
\includegraphics[width=1\linewidth]{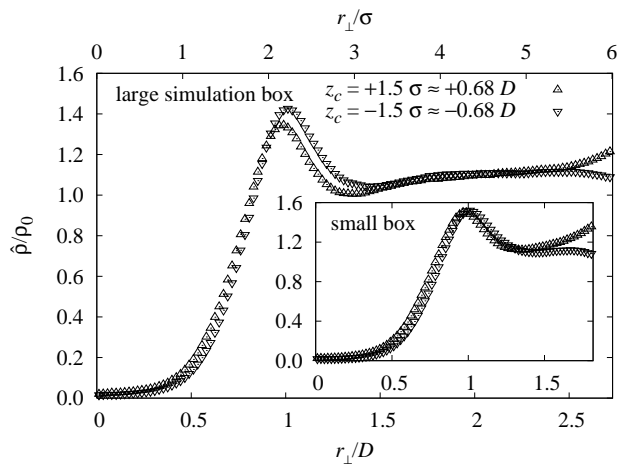}
\end{center}
\caption{
  The plot represents $\hat{\rho}(r_\perp,z_c)$, see
  Eq.~(\protect\ref{eq:radialrho}), in a Stokes flow
  ($R^*=0.77$) with $c^\ast=22$ upstream (at $z_c=1.5\sigma$,
  triangles up) and
  downstream (at $z_c=-1.5\sigma$ triangles down) of the colloid,
  as measured in BD simulations. The inset shows the results for a
  smaller simulation box. }
\label{fig:finite}
\end{figure}

We notice a discrepancy between the density profiles measured in the BD
simulation and those calculated numerically in the dDFT. We
attribute this to two finite--size effects in the simulation
which have been confirmed by performing BD simulations in a smaller
box at the same polymer number density ($L_x=L_y=8\,\sigma$, 
$L_z=16\,\sigma$, 
$N=1024$)\/.
First, the numerical solution of the dDFT for a Stokes flow with
$c^\ast=22$ exhibits a long ($\approx 15\,\sigma$) tail of slight
polymer depletion ($\rho \approx 0.98\,\rho_0$) behind the colloidal
particle. The tail is longer than the length of the BD simulation box 
in $z$-direction and, as a 
consequence of the periodic boundary conditions, the inflowing density 
far ahead of the particle is smaller than $\rho_0$\/.
This screening effect is very noticeable when the flow is approximated
as uniform because the depletion of polymers behind the colloidal
particle is very large, see Fig.~\ref{fig:BDvsdDFT}(b)\/.  But in the
case of the Stokes flow, this effect seems to less
important compared to the second effect:
%
the discontinuity of the normal component of the flow
field at the lateral boundaries of the simulation box leads to a
non-vanishing divergence of the flow there (we remind that the flow
described by Eq.~(\ref{eq:sphereflow}) has $\nabla\cdot\vct{u} =
0$)\/.
Fig.~\ref{fig:finite} represents the density profile
averaged over thin cylindrical shells of height $\sigma$ and radial
thickness $\Delta r=0.05\,\sigma$ 
coaxial with the $z$-axis, i.e.,
\begin{equation}
  \label{eq:radialrho}
  \hat{\rho}(r_\perp, z_c) := 
  \frac{1}{\sigma \left(r_\perp+\frac{\Delta r}{2}\right) \Delta r}
  \int\limits_{z_c-\frac{\sigma}{2}}^{z_c+\frac{\sigma}{2}}\!\! dz'
  \int\limits_{r_\perp}^{r_\perp+\Delta r}\!\! dr'_\perp
  r'_\perp \, \rho(r'_\perp, z') .
\end{equation}
The periodic boundary conditions imply that $\nabla\cdot\vct{u}<0$
effectively at the side boundaries located upstream, where
therefore the density is enhanced: Even though we expect the
density to decrease towards $\rho_0$ as the radial
distance $r_\perp$ to the colloid increases we find instead an increase of
the density at the boundary of the simulation box.
At the side boundaries located
downstream, on the other hand, $\nabla\cdot\vct{u}>0$ and the region
near the boundary of the simulation box becomes depleted of
polymers.
As expected this effect is enhanced for reduced box size, see the
inset in Fig.~\ref{fig:finite}\/.
The comparison of the BD results with the dDFT calculation in
Fig.~\ref{fig:BDvsdDFT} indicates that the overall consequence of these
effects is a density enhancement near the colloidal particle.
In view of these important finite--size effects, we cannot quantify the
validity of the approximation given by Eq.~(\ref{eq:dDFT}) for a
realistic 
flow. When there is only uniform flow
($R^\ast=0$), however, the finite--size effects are much 
less pronounced and we
find a good agreement between BD simulations and dDFT calculations,
in concordance with Ref.~\cite{penna03b}\/.

\subsection{Drag force on colloids}
\label{sec:drag}

We have also measured the force in the
$z$-direction exerted by the polymers on the particle. This force is
additional to the Stokes drag force $F_{\text{Stokes}}$ 
exerted by the flowing solvent.
If $\Gamma$ in Eq.~(\ref{eq:brownian}) is assumed to be given by the
Stokes-Einstein relation for spherical polymers of diameter
$\sigma$, the Stokes drag for a colloid of radius $R$ in the same
solvent is given by
$F_{\text{Stokes}} = 2\,c^\ast\,(R/\sigma)\,(T/D)$\/. For
example, for the colloid radius $R^\ast=0.77$ considered in the
BD simulations this gives $F_{\text{Stokes}}  = 3.4\,
c^\ast \,T/D$\/.

In the BD simulations, the force excerted on the colloid by the
polymers can be measured directly. In the case of ideal particles
discussed in Sec.~\ref{sec:ideal} we use the ideal gas law
$p=T\,\rho$ in order to calculate the local pressure on the colloid
surface. Integrating the local pressure over the surface yields the
force on the colloid.

Table~\ref{tab:force} collects the
mean force for different types of flow
($R^\ast=0$ for uniform flow and $R^\ast=0.77$ for Stokes
flow) and
velocities. The results confirm the necessity to take the solvent
flow into account: 
the mean force in the case of Stokes flow is markedly smaller 
(at most of the order of
$F_{\text{Stokes}}$) than in the case of a homogeneous flow and the
dependence on $c^\ast$ is milder. This can be understood in terms
of the reduction of the bow wave effect in the density profile around the
particle by the Stokes flow in the solvent which advects the
polymers. The forces in the BD simulations are of the same order of
magnitude as in the ideal case, but in the case of the Stokes flow
they have a weaker dependence on $c^\ast$\/. 

Because the sizes involved are of the order of the microscopic length
$D$, the variance of the force measured in the Brownian
dynamics is relatively large. However, we find that it
is not affected by the flow type and velocity and it coincides with
the variance of $F_z$ we have measured in the equilibrium state
($c^\ast=0$)\/.
For comparison, Tab.~\ref{tab:finite} collects the force measured in
the BD simulation in the smaller box. The increased force is
consistent with the density enhancement near the particle caused by
the finite--size effects. We also attribute the increase of the
fluctuations to these effects.

\begin{table}[t]
  \begin{ruledtabular}
    \begin{tabular}{ccccc}
      Type of flow & $c^\ast$ &
      $|\langle F_z \rangle|$ &
		ideal $F_z$ &
		$F_{\text{Stokes}}$
		\\\hline
      uniform & 2.2 & 42.5 &  41.9 & \\
      Stokes & 2.2 & 6.38 &  3.14 &  7.48\\
      uniform & 22 & 215 &  290 & \\
      Stokes & 22 & 10.3 &  18.0 & 74.8\\
    \end{tabular}
  \end{ruledtabular}
  \caption{
    Mean force 
	 exerted by the polymers measured 
    in the BD simulations for different types of flow ($R^\ast=0$ for
	 the uniform flow, and $R^\ast=0.77$ for the Stokes flow),
	 compared to the force exerted by ideal polymers
	 and to the Stokes friction of the colloid. 
	 The forces are given in units of $T/D$.}
  \label{tab:force}
\end{table}

\begin{table}[t]
  \begin{ruledtabular}
    \begin{tabular}{cccc}
      simulation box size& $c^\ast$ &
      $|\langle F_z \rangle|$ &
      $\sqrt{\langle F_z^2 \rangle - \langle F_z \rangle^2}$
		\\\hline
      small & 2.2 & 11 & 63.4 \\
      large & 2.2 & 6.4 & 103 \\
      small & 22 & 19 & 65.6 \\
      large & 22 & 10 & 103 \\
    \end{tabular}
  \end{ruledtabular}
  \caption{
    Mean and variance of the force exerted by the polymers in a
	 Stokes flow measured in BD simulations of different boxsizes
	 (see Sec.~\protect\ref{sec:gauss})\/.
	 The forces are given in units of $T/D$\/. }
  \label{tab:finite}
\end{table}


\section{Conclusions} \label{sec:conclusions}

We have proposed a dynamic density functional theory (dDFT),
Eq.~(\ref{eq:dDFT}), for interacting Brownian particles in a flowing
solvent under the assumption that detailed balance holds (which
requires, in particular, a curl-free flow).
We get the same equation as already derived in \cite{marconi99} but
with the partial time derivative replaced by the total (material) time
derivative. The whole effect of the flow field can be summarized into
a modified external potential, allowing application of the whole
machinery of equilibrium statistical mechanics. Thus, we are able to
find an H-theorem for a modified free energy. 

In this paper we include the displacement of the solvent 
by the colloid, but the hydrodynamic interactions between the colloid
and the solute particles as well as the hydrodynamic interactions
among the solute particles were not taken into account. While the
latter is a highly non-trivial and still open problem, the first
can be included in a straightforward manner by replacing the
mobility $\Gamma$ by a space dependent and symmetric
mobility tensor $\tens{\Gamma}(\vct{r}_i)$ in
Eqs.~(\ref{eq:brownian}) and (\ref{eq:noisecorrelator})\/. Thereby
the noise becomes multiplicative and the appropriate calculus has to
be considered such that it leads to the Fokker-Planck equation
(\ref{eq:fokkerplanck}) with $\Gamma$ replaced by
$\tens{\Gamma}(\vct{r}_i)$\/. Then the equilibrium distribution is
not changed. For spherical particles in the vicinity of planar
walls the mobility tensor can be calculated in the limit of large
distances \cite{happel65}\/. This result has been extended to
surfaces with a partial slip boundary condition 
in \cite{lauga05b}\/. Due to the translational symmetry of
the system, $\tens{\Gamma}$ is diagonal. While the mobility
perpendicular to the wall increases with distance, the distance
dependence of the mobility
parallel to the wall depends on the slip condition. For no-slip it
increases while for total slip it decreases with the distance to
the wall. The hydrodynamic interaction between two spheres has
been calculated, e.g., in the Rotne-Prager approximation
\cite{jeffrey84,dhont97}\/. The derivation of
Eq.~(\ref{eq:dDFT}) essentially carries through with the only
exception that the quotient $\vct{u}(\vct{r_i})/\Gamma$ in
Eq.~(\ref{eq:fokkerplanck}) has to be
replaced by a field $\vct{\tilde{u}}(\vct{r}_i)$ with
$\tens{\Gamma}(\vct{r}_i)\cdot
\vct{\tilde{u}}(\vct{r_i})=\vct{u}(\vct{r_i})$\/. In order to
absorb the flow field into a modified external potential,
$\vct{\tilde{u}}$ (and not only $\vct{u}(\vct{r_i})$) has to be curl
free with $\vct{\tilde{u}}=-\grad\Phi$\/. Instead of 
Eq.~(\ref{eq:dDFT}) we then get
\begin{equation}
\frac{\partial \rho}{\partial t}+\grad\cdot(\rho\,\vct{u})=
\grad\cdot\left(\rho\,\tens{\Gamma}(\vct{r})\cdot\grad\frac{\delta \HF[\rho]}{\delta\rho}\right).
\end{equation}

In Ref.~\cite{penna03b}, the polymer distribution was studied in a
polymer solution flowing uniformly through a spherical particle which
is hard only for the polymers. In spite of the violation of detailed
balance by the boundary conditions, the comparison between simulations
and the numerical solution of the proposed dDFT was good.  In this
paper we have considered the more realistic case of a Stokes
flow~(\ref{eq:sphereflow}) around the particle. 
The exact solution of the ideal case (no polymer--polymer
interaction), the numerical solution of the interacting case as
well as the corresponding Brownian dynamics simulations
evidence all the dramatic effect by advection on
the properties of the stationary solution. We conclude that the
approximation of uniform flow, as employed in
Refs.~\cite{penna03b,dzubiella03b,krueger07}, is quantitatively bad. 
%
We have found discrepancies in the density distribution of polymers as
measured in the simulations and as computed numerically in the
framework of the dDFT. However, the discrepancies could be
rationalized in terms of finite--size effects in the simulations due
to the slow decay of the Stokes flow far from the obstacle. Thus,
although a quantitative check of the validity of the
approximations leading to the dDFT in Eq.~(\ref{eq:dDFT}) and of
its validity for non-potential flows was not possible
the results are encouraging.

\begin{acknowledgments}
  The authors thank S.~Dietrich for financial support and fruitful
  discussions. 
  A.~D. acknowledges financial support from the Junta
  de Andaluc{\'\i}a (Spain) through the program ``Retorno de
  Investigadores''.
  M.~R. acknowledges funding by the Deutsche
  Forschungsgemeinschaft within the
  priority program SPP~1164 ``Micro- and Nanofluidics'' under grant
  number RA~1061/2-1.
\end{acknowledgments}






\end{document}